\documentclass[letter,superscriptaddress]{revtex4}
\usepackage{epsfig}
\usepackage{graphicx}

\begin{document}
\title{The multiferroic phase of DyFeO$_{3}$:\\
an ab--initio study.}

\author{Alessandro Stroppa}
\affiliation{CNR-SPIN, L'Aquila, Italy}
\author{Martijn Marsman} 
\affiliation{Faculty of Physics,  and Center for
Computational Materials Science, University of Vienna, Sensengasse
8/12, A-1090 Wien, Austria}
\author{Georg Kresse}
\affiliation{Faculty of Physics,  and Center for 
Computational Materials Science, University of Vienna, Sensengasse
8/12, A-1090 Wien, Austria}
\author{Silvia Picozzi}
\affiliation{CNR-SPIN, L' Aquila, Italy }

\begin{abstract}
By performing  accurate ab-initio density functional theory calculations,
we study the role of  $4f$ electrons in stabilizing the magnetic-field-induced 
ferroelectric state of DyFeO$_{3}$. We confirm  that the ferroelectric polarization  is  driven by an exchange-strictive mechanism, working between adjacent spin-polarized Fe
and Dy layers, as suggested by Y. Tokunaga [Phys. Rev. Lett, \textbf{101}, 097205 (2008)].  A careful electronic structure analysis suggests that 
coupling between  Dy  and Fe spin sublattices is mediated by Dy-$d$ and O-$2p$ hybridization. Our results are  robust with respect to the different computational schemes used for $d$ and $f$ localized states, such as the DFT+$U$ method, the Heyd-Scuseria-Ernzerhof (HSE) hybrid functional and  the GW approach. Our findings  indicate that the interaction between the $f$ and $d$ sublattice might be used to tailor ferroelectric and magnetic properties of multiferroic compounds. 
\end{abstract}


\maketitle
\section{Introduction}
\label{introduction}\label{intro}
Multiferroics (MFs) are compounds where 
long-range magnetic and dipolar orders coexist\cite{Aizu}.
There is plenty of fascinating physics in these materials,
due to the strong entanglement of spin-charge-orbital degrees of freedom\cite{Scott.Nature,Renaissance} and a  great potential for technological 
applications has already been recognized\cite{Applications1,Applications2,Applications3}.
The coupling between magnetism and ferroelectricity can be used for sensing applications, but also for memory devices where data is typically stored as magnetic information and read out 
electrically. Recently, several manganese and iron oxides have been shown 
to possess strong coupling;  however, ferroelectricity in these materials
is rather weak and only the electrical polarization can be switched by a magnetic field
 (but not viceversa, a limitation for many applications). Within this framework, 
DyFeO$_{3}$  is a very interesting compound\cite{Acta1,Acta2} because
it shows large ferroelectric polarization combined with a strong magnetoelectric 
coupling. Furthermore, the ferroelectric polarization in DyFeO$_{3}$ is induced 
by the peculiar magnetic structure, obtained upon applying an external magnetic field.

 DyFeO$_{3}$ belongs to the   the class of    perovkite oxides, 
such as RMO$_{3}$ (where R is a rare-earth ion and M is a 
transition metal atom).  From the computational point of view, their description poses serious problems.  These compounds  are often
(strongly) correlated materials, 
involving $d-$ and $f-$ electronic charge with
significant spatial localization.
First-principles density functional theory (DFT) calculations in the 
most commonly applied  local density approximation (LDA) or generalized gradient approximation (GGA)
have to face well
known deficiencies: 
the non-locality of the screened exchange interaction is not  well
taken into account and the electrostatic self-interaction is not entirely
compensated\cite{NatMat}.
Since semilocal functionals tend to delocalize $f$-states,
the $f$-electrons are often kept ``frozen'' in the core, and the origin of multiferroicity is generally attributed to the spin-charge-orbital degrees of freedom of the M sublattice. Although this ``standard'' approach  helped  in clarifying many mechanisms leading to ferroelectricity in MFs\cite{twist},
  the influence of 
$f$ electrons on multiferroicity  has not been extensively investigated yet 
by \textit{ab--initio} 
 calculations, despite  several
experiments clearly point out  $f$-electrons to play an important role in MF properties of RMO$_{3}$ compounds\cite{expf1,expf2}. This is especially so for DyFeO$_{3}$, where the 
electric polarization  
   appears at a very low temperature, 
corresponding to ordering of Dy spins\cite{Prediction,Tokura.DFO,Nature.GdFeO3}. Recall that
 Dy is expected to have a $3+$ oxidation state, \textit{i.e.} a $f^{9}$ configuration\cite{f9}.

In this study, stimulated by the recent experimental demonstration of a
magnetic-field-induced ferroelectric (FE)  phase of DyFeO$_{3}$ (DFO)\cite{Prediction,Tokura.DFO,Nature.GdFeO3},  
we performed ab-initio simulations in order to show the key role played by $f$ electrons  in stabilizing spin-driven ferroelectricity. Our calculations support the experimental study of Ref.\ \onlinecite{Tokura.DFO,Nature.GdFeO3}
 and, at the same time, quantify the polarization and shed light on the microscopic mechanism 
(\textit{i.e.} based on electronic structure analysis) of the origin of  
ferroelectricity.  To the best of our knowledge, this is the first \textit{ab--initio} study of the above-mentioned effect. Our simulations were mainly carried out within a DFT+$U$ approach for localized electrons. In addition, we used the  
Heyd-Scuseria-Ernzerhof (HSE) screened hybrid functional\cite{hse1,hse2}, which has been shown 
to improve the description of $d$- and $f$-electron systems\cite{hse3,hse4,hse5} over LDA or GGA. 
Finally, our calculations were benchmarked by single shot GW using HSE wave functions ($G_{0}W_{0}@HSE$); this treatment is expected to give a very accurate description of the electronic ground state\cite{GWHSE}. 
\begin{figure} 
\centering
\includegraphics[width=0.45\textwidth,angle=0,clip=true]{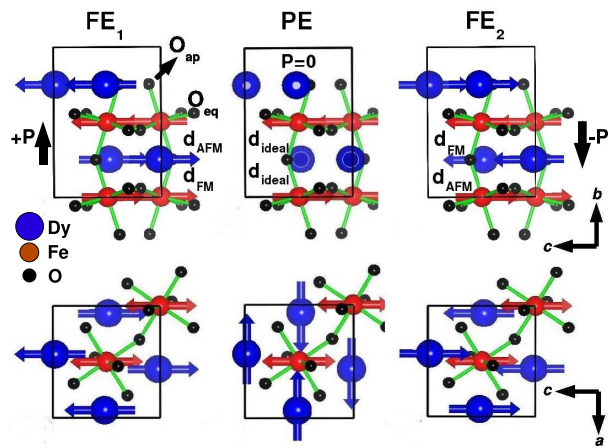} 
\caption{(Color online) Ball-and-stick model of DFO in the  PE and FE states (FE$_{1}$ and FE$_{2}$). Upper part: side views; bottom part: top view along the $b$ axis. 
Blue (large), red (medium) and black (small) spheres show Fe, Dy, and O atoms, respectively. 
O$_{eq}$, O$_{ap}$ refer to \textit{equatorial} and \textit{apical} oxygens. 
Arrows represent spin moments.}\label{DFO}
\end{figure}

\section{Computational details}
Calculations were performed using the projector augmented-wave
(PAW) method\cite{paw1,paw2} with the Perdew-Burke-Ernzerhof
(PBE) GGA functional\cite{pbe}. We used DFT+$U$ within Dudarev's approach\cite{ldau} using $U_{\rm eff}=U-J$=3 eV and 4 eV for Fe-$d$ and Dy-$f$ states, respectively. The energy cutoff was set to 400 eV and a 4$\times$2$\times$4 Monkhorst-Pack grid of $k$-points was used. We treated  the Dy $f$ electrons both as  valence and as core states\cite{vaspmanual}. The Berry phase approach\cite{berry1,berry2} was used to calculate the macroscopic polarization $P$. Non collinear magnetism was treated in accordance with Ref.\ \cite{NC}. Spin-orbit coupling (SOC) was included for the end--point states  of the adiabatic path, 
{\emph i.e.} paraelectric (PE)  
and ferroelectric (FE)  DyFeO$_{3}$,
see below. For hybrid functionals, we used the HSE functional, recently implemented in VASP\cite{krukau}. $GW$ calculations\cite{GW1,GW2} were performed on top of the HSE ground state\cite{GWHSE}.
The experimental lattice constants for orthorhombic DFO were used (space group $Pnma$, with $a$=5.596 \AA, $b$=7.629 \AA, $c$=5.301\ \AA)\cite{Acta1}. Starting from experimental atomic positions, we performed atomic relaxations until residual Hellman-Feynman forces were $<$0.01 eV/\AA.

In figure \ \ref{DFO}, we show the two  FE states of DFO, FE$_{1}$ and FE$_{2}$, with opposite polarization. These represent a simplified version of the complex experimental magnetic-field-induced polar states (cfr figure\ 1 (c) and (d) of Ref.\ \onlinecite{Tokura.DFO}), obtained by neglecting the $x$ and $z$ components of Dy and Fe spins.\footnote{In our notation, the $b$ axis  correspond to  the $c$  axis in Ref.\ \onlinecite{Tokura.DFO}}.
In both, the FE$_{1}$ and the FE$_{2}$ cases, the Fe and Dy  sublattices show an A-type magnetic structure, \textit{i.e.} ferromagnetic (FM) and antiferromagnetic (AFM) intralayer and interlayer coupling, respectively, with the spins pointing along the in--plane $c$ axis. In FE$_{1}$, 
the stacking of the spins  along  the out-of-plane $b$ axis 
is \textbf{$\downarrow$}Fe-$\downarrow$Dy-$\uparrow$Fe-$\uparrow$Dy;
the FE$_{2}$ state is obtained from the FE$_{1}$ state 
by rotating Dy spins by 180$^\circ$, so that the spin stacking  is $\downarrow$Fe-$\uparrow$Dy-$\uparrow$Fe-$\downarrow$Dy. In the PE state, the atoms are arranged according to  the   $Pnma$ ($D_{2h}$) space group and
the Dy spins are  still intra-layer FM coupled, but rotated with respect to the Fe spins by 90$^\circ$.  
In  FE$_{1}$, the spins of a Fe layer become parallel to the moments on one of the nearest-neighbor Dy layers  and antiparallel to the other:   Dy layers should then displace cooperatively towards Fe layers with parallel spins via exchange striction, giving rise to alternating short-long-short-long interlayer Dy-Fe distances. Accordingly, a polarization $P$ along the $b$ axis should be generated. In the FE$_{2}$ state, the flip of Dy   spins would  cause a reversal of $P$. 
Finally, in the PE  state, each 
Fe sheet is sandwiched  by layers with $\perp$ Dy spins, and no interlayer dimerization is expected. This microscopic mechanism, which involves frustrated interactions between rare earth and transition metal ions
and its optimization  by exchange striction, was proposed in Refs.\ \onlinecite{Tokura.DFO,Nature.GdFeO3}.

\section{Switchable ferroelectric states}
Our calculations confirm 
the above interpretation. First of all, the total energies of FE$_{1}$ and FE$_{2}$ are degenerate.\footnote{The two values differ by less than 1 meV/cell} Furthermore, from the symmetry point-of-view, the rotation of Dy spins 
from PE to FE$_{1}$ or to FE$_{2}$  
causes a symmetry lowering from the non-polar 
space group 62 (D$_{2h}$) 
to a polar space group 33 (C$_{2v}$), giving rise 
to a polarization along the $b$ axis.  Indeed, the distortions lead to an alternate 
short-long-short-long interlayer distance, 
  with $d_{FM}=$ 1.898\ \AA\ 
and $d_{AFM}=$ 1.910\ \AA, while $d_{ideal}$ =  1.904\ \AA\ in the unrelaxed non-spin 
polarized case\footnote{Further relaxations including  SOC give 
$d_{FM}=$ 1.899\ \AA\ and $d_{AFM}$=1.909\ \AA, whereas HSE gives: 
$d_{FM}=$ 1.899 \AA\ and $d_{AFM}=$1.910 \AA.}. Clearly, the Dy-\textit{$f$ electrons play a key role} in stabilizing the FE state. In order to prove this, we  froze the $f$ electrons in the core
and we calculated the electronic ground state using the previously relaxed FE$_{1}$ structure.  First, the electronic contribution to the polarization, $P_{ele}$ vanishes;  then, by allowing the ions to relax, the ionic contribution,  
$P_{ionic}$, becomes negligible as well, and
the final crystal structure is non-polar.  
In summary, when treating the $f$ electrons as valence states, the PE state becomes unstable, the D$_{2h}$  point group symmetry is spontaneously broken to C$_{2v}$ and the system evolves towards a stable and polar state. If the $f$ electrons are removed from the valence and frozen in the core,
 the PE state remains stable. This unambiguously confirms that $f$ states are a necessary ingredient for ferroelectricity in DyFeO$_{3}$.
\begin{figure}
\centering
\includegraphics[totalheight=0.33\textheight,angle=0,clip=true]{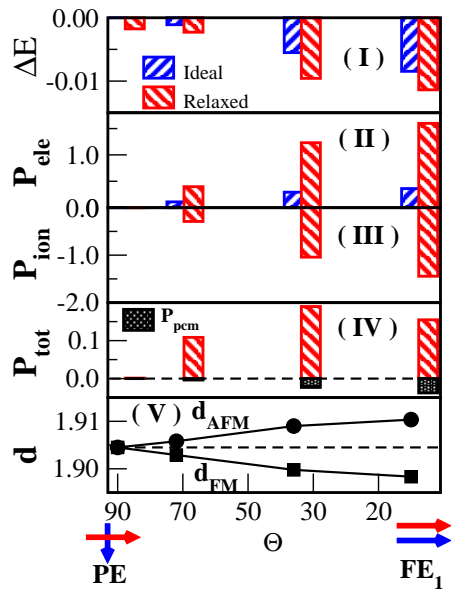}

\caption{(Color online) Adiabatic path connecting the PE and the FE$_{1}$ phases through step-wise rotation of the Dy spin direction direction from 90 to 0$^\circ$.
See text for details. Units are (eV/cell) for $\Delta$E, $\mu$Ccm$^{-2}$ for FE polarization, \AA\ for interlayer distance ($d$).}\label{Path}
\end{figure}
\section{Discussion}
\subsection{Adiabatic path} 
To shed light on the onset of ferroelectricity,
we construct an adiabatic path by  progressively rotating  the Dy spins from 90 to 0 degree, \textit{i.e.} from 
the PE to FE$_{1}$ state. The results are summarized in figure\ \ref{Path}: 
panel (I) shows the energy difference between 
the FE$_{1}$ and the PE phases, evaluated at the ideal centrosymmetric 
(CS) ionic structure (in blue) and relaxed configuration
(in red); in (II), (III), (IV) we show the electronic, ionic and total FE polarization, respectively, evaluated at the ideal (blue) and relaxed (red) ionic structure; 
in (IV) the Point-Charge-Model (PCM) estimate of the polarization is also reported (in black); in (V) the change in the interlayer distances along the path (see also Fig.\ \ref{DFO}) is shown.
Panel (I) clearly shows that ferroelectricity in DFO is magnetically induced. 
In fact, already for the CS structure, the spin rotations 
give rise to an energy gain (blue bars),
further enhanced by ionic relaxations (red bars).
Note that the relaxation of  the electronic degrees of freedoms accounts for most of the total stabilization energy (cfr blue and red bars). The energy gain increases from left to right in panel (I), \textit{i.e.} towards a collinear configuration. 
In parallel, $P_{ele}$ also increases from left to right, even in the ideal CS structure (blue), as expected for magnetically induced MFs\cite{Silvia}; ionic relaxations further increase $P_{ele}$ [see panel (II)]. From panel (III), we see that 
$P_{ion}$ is opposite to $P_{ele}$ and of the same order of magnitude. However, $P_{ion}$ does \textit{not} fully compensate $P_{ele}$  [see panel (IV)], giving rise to a total polarization $P_{tot}$ of $\sim$ 0.20 $\mu$Ccm$^{-2}$ for collinear spins. The inclusion of SOC confirms $P_{tot}$, \textit{i.e.}  0.18 $\mu$Ccm$^{-2}$. Notably, this value is in good agreement with the estimate given in Ref.\ \cite{Tokura.DFO}. HSE also confirms the value of  FE polarization.
Furthermore, note that $P_{pcm}$ is not only opposite to $P_{tot}$, but also smaller in absolute value than $P_{tot}$. This confirms that  electronic degrees of freedom trigger the FE transition.
Finally, in panel (V), the evolution of the  interlayer distances along the polar \textit{b} axis clearly shows that the Dy and Fe layers are coupled, giving rise to spin-driven interlayer dimerization.\\
\begin{figure}
\centering
\includegraphics[width=0.45\textwidth,angle=0,clip=true]{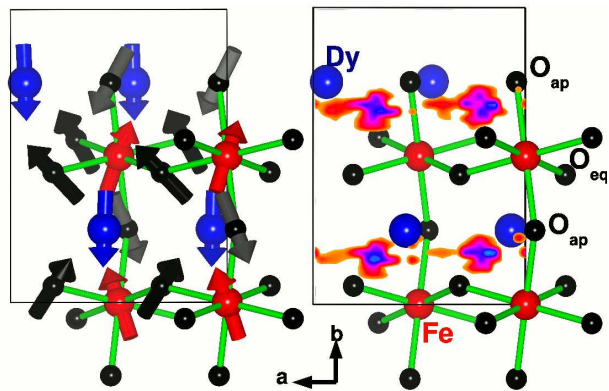}
\caption{(Color online) Left: Atomic displacements (arrows) between the CS  
and the FE$_{1}$ structure. Spins are not shown for clarity. Right: Projection of positive 
DELF($\vec{r}$) isosurface in the
$ab$ plane of FE$_{1}$ structure. See text for details.}\label{Displ}
\end{figure}

\subsection{Ionic relaxations and origin of the ferroelectricity}
In the PE phase Fe, Dy, O$_{ap}$ (apical oxygens), O$_{eq}$ (equatorial oxygens) occupy the $4a$, $4c$, $4c$, and $8d$ Wyckoff positions (WPs), respectively. When the symmetry is lowered to C$_{2v}$, Dy and O$_{ap}$ change their site symmetry to $4a$ and the
O$_{eq}$ become \textit{inequivalent} ($8d\rightarrow4a+4a$). This is readily explained by considering the local spin configuration around O$_{eq}$s: when the latter are sandwiched by  FM coupled
Fe and Dy layers, they have  two $\uparrow$Fe and two $\uparrow$Dy atoms as nearest neighbors (these will be labelled O$_{eq}^{\uparrow,\uparrow}$); 
when sandwiched by Fe and Dy layers AFM coupled, they have   two $\uparrow$Fe and two $\downarrow$Dy atoms as nearest neighbors (labelled as   O$_{eq}^{\uparrow,\downarrow}$).  First of all, all  O$_{eq}$s carry a spin-induced moment parallel to the neighboring  Fe atom. Furthermore, by 
imposing the FE$_{1}$ spin configuration 
on top of the CS ionic structure, O$_{eq}^{\uparrow,\uparrow}$ and  O$_{eq}^{\uparrow,\downarrow}$ become  \textit{inequivalent}: the O$_{eq}^{\uparrow,\uparrow}$ has $\pm$0.194 $\mu_{B}$ and  the O$_{eq}^{\uparrow,\downarrow}$ has $\pm$0.207 $\mu_{B}$ 
as induced spin moment. 
To rule out any numerical artifact on this small difference, we impose the PE spin configuration on top of the CS ionic structure. In this case, all O$_{eq}$s carry induced spin moments of exactly the same magnitude. 
In passing, we note that all oxygens remain equivalent when freezing the Dy-$f$ electrons in the core.
Furthermore, we performed an analysis of the 
 symmetry breaking distortions\cite{Amplimode,Isodisplace}.
The mode decomposition confirms that a FE mode is involved, called $GM4-$. 
In Fig.\ \ref{Displ} we show the pattern of atomic displacements with respect to the CS structure.
The inequivalency of O$_{eq}$ is subtle: O$_{eq}^{\uparrow,\uparrow}$ (O$_{eq}^{\uparrow,\downarrow}$) move in such a way to \textit{decrease}  (\textit{increase}) the distance to its neighbor Dy atom. For O$_{eq}^{\uparrow,\uparrow}$, d$_{Dy-O}$ is 2.478 \AA; for  O$_{eq}^{\uparrow,\downarrow}$, d$_{Dy-O}$ is 2.496 \AA\ (the same distance in the PE phase is 2.487 \AA), suggesting that
a weak bonding interaction is active between the FM layers, leading to a polarization along $b$.
A useful tool  for studying tiny differences in bonding interaction
in solid state systems is the electron localization function (ELF)\cite{Savin1,Savin2}.
The ELF values lie by definition between zero and one. Values
are close to 1, if in the vicinity of one electron no
other electron with the same spin may be found, for instance as occurs
in \textit{bonding pairs}.
 Here, we consider the\textit{ difference} in ELF (DELF)
between the situation when $f$ electrons are in the valence and when they are frozen in the core, for the same ionic configuration, \textit{i.e.} 
DELF($\vec{r}$)=ELF$_{f_{val}}$($\vec{r}$)-ELF$_{f_{core}}$($\vec{r}$).
The physical interpretation is as follows: positive values 
of DELF show up in regions where the electron localization is higher, \textit{i.e.} the bonding between FM layers is strengthened. In Fig.\ \ref{Displ} (right part) we show a \textit{positive} isosurface  of DELF 
projected into the $ab$ plane. Clearly, it is  
mainly localized between FM layers and, more specifically, in the region between Dy and O$_{eq}^{\uparrow,\uparrow}$. 
This points to a bonding interaction between FM layers mediated by O$_{eq}^{\uparrow,\uparrow}$. 

\subsection{Electronic structure fingerprint}
\begin{figure}
\centering
\includegraphics[totalheight=0.3\textheight,angle=0,clip=true]{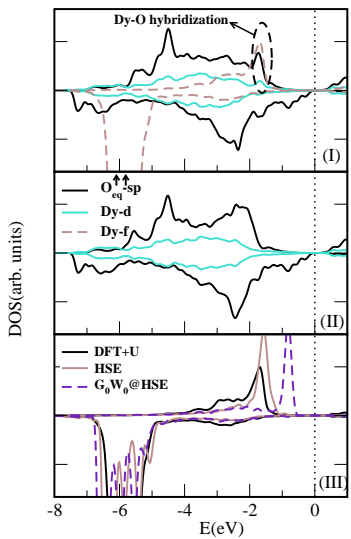}
\caption{(Color online) DOS for oxygen and Dy atoms for $f$ as valence [panel (I)]  or as core 
states [panel (II)]. Panel (III) shows the $f$-DOS as calculated by DFT+$U$, HSE and G$_{0}$W$_{0}$@HSE. Vertical dotted lines refer to the Fermi level. Positive (negative) DOS values refer to 
minority (majority) states. See text for details.}\label{dos}
\end{figure}

A careful inspection of the orbital-decomposed magnetic moments reveals that the Dy-$5d$ states are polarized \textit{only} if the $4f$ states are in the valence: the $4f$ states couple locally to the  $5d$ spin moments  by intra-atomic $4f$-$5d$ exchange interaction\cite{f_d}. The $5d$ states are much more extended than $4f$ electrons, suggesting that the  glue that finally couples FM Fe and Dy layers is the interatomic interaction between the Fe $3d$ states and the Dy $5d$ states  mediated by the oxygen $p$ states.  
In Fig.\ \ref{dos} we show the local Density of States (DOS) at O$_{eq}^{\uparrow,\uparrow}$ atoms and the Dy-$d$ and Dy-$f$ states. 
In panel (I), a clear interaction between $d$ and $f$ states of Dy and oxygen states is visible (cfr dotted ellipse). Note that the same interaction involves the  Fe $3d$ states as well (not shown in Fig.\ \ref{dos}).
Remarkably, the interaction disappears when freezing the $f$ states in the core (see panel (II)), for which the intra-atomic Dy-$f$-$d$ and interatomic O-$p$ hybridization disappears. 
Panel (III) shows the Dy-$f$ DOS calculated using DFT+$U$, HSE and G$_{0}$W$_{0}$@HSE calculations. In the (relevant) occupied manifold, the chosen $U$ nicely fits the HSE DOS, which, in turn, is rather close to the  G$_{0}$W$_{0}$ calculations. The corrections beyond a mere DFT+$U$ approach show up in the unoccupied states by opening the band gap; however, this does not change our conclusions, as far as the polarization is concerned.

The fact that the Dy and Fe interaction is mediated by O$_{eq}^{\uparrow,\uparrow}$-$sp$ states suggests possible routes to tailor 
the ferroelectric polarization. For instance,   compressive or tensile strain along the $b$ axis should increase or decrease the tilting of the octahedra, favoring or disfavoring the interaction via the intermediate  O$_{eq}^{\uparrow,\uparrow}$
states, \textit{i.e.} enhancing or reducing the ferroelectric polarization.

\section{Conclusions} 
Several   results emerge from our study: 
{\em i)} 
the FE state of DFO is driven by
\textit{exchange striction}, confirming the qualitative explanation given in Refs.\onlinecite{Tokura.DFO,Nature.GdFeO3}; {\em ii)} two degenerate and switchable polar
 states exist characterized by a sign-reversal of the FE polarization ($\pm P$)
 and linked by  an adiabatic path (connecting the  two ferroelectric (FE) states FE$_{1}$ and FE$_{2}$ to the same paraelectric reference structure) can be obtained through 
 a relative rotation of the direction of Dy spins (with respect to Fe spins); 
 {\em iii)} the coupling between  Dy  and Fe spin sublattices is mediated by Dy-$d$ and O-$2p$ states; 
 {\em iv)} the estimated FE polarization is in agreement with experiments; {\em (v)} by    
freezing the $f$ states in the core instead of  relaxing them in the valence, we confirm  the crucial
role played by $f$ electrons in establishing the spin-driven ferroelectricity.
More generally, our study suggests that $f$ electrons might play an important role in the
ferroelectric properties of other RMO$_{3}$ compounds (where $4f$ electrons  are often 
neglected in theoretical calculations); {\em (v)} last, but not least, 
our electronic structure analysis suggests possible routes to tailor 
the ferroelectric polarization, owing to the strong Dy-Fe coupling via intermediate equatorial oxygens. Further study is in progress to confirm 
this expectation.


\section{Acknowledgments} 
The research leading to these results has received funding from the European Research Council under the EuropeanCommunity,  7th Framework Programme - FP7 (2007-2013)/ERC Grant Agreement n. 203523. 
Computational support by Caspur Supercomputing Center in Roma is gratefully acknowledged. Figures  done 
using the VESTA package.\cite{vesta} A.S. and S.P. thank D. Khomskii and N. Spaldin
for useful insights. A.S. thanks  G. Giovannetti and K. Yamauchi 
for interesting discussions.  M.M. and G.K. acknowledge support by the Austrian science fund (FWF).

\end{document}